\documentclass[10pt, conference, letterpaper]{IEEEtran}
\IEEEoverridecommandlockouts
\usepackage{graphicx}
\usepackage{booktabs}
\usepackage{amssymb}
\usepackage{tabularx}
\usepackage{multirow}
\usepackage{tablefootnote}
\graphicspath{ {./images/} }
\usepackage[hidelinks]{hyperref}
\usepackage{enumerate}
\usepackage[utf8]{inputenc}
\usepackage{pgfplots}
\DeclareUnicodeCharacter{2212}{−}
\usepgfplotslibrary{groupplots,dateplot}
\usetikzlibrary{patterns,shapes.arrows}
\pgfplotsset{compat=newest}
\usepackage{amsmath}
\usepackage[mathscr]{euscript}
\usepackage{caption}
\usepackage{subcaption}
\usepackage{numprint}
\usepackage{enumitem}
\usepackage{comment}
\npthousandsep{\,}
\usepgflibrary{patterns} % LATEX and plain TEX and pure pgf
\usepgflibrary[patterns] % ConTEXt and pure pgf
\usetikzlibrary{patterns} % LATEX and plain TEX when using TikZ
\usetikzlibrary[patterns] % ConTEXt when using TikZ
\usepgflibrary{patterns.meta} % LATEX and plain TEX and pure pgf
\usepgflibrary[patterns.meta] % ConTEXt and pure pgf
\usetikzlibrary{patterns.meta} % LATEX and plain TEX when using TikZ
\usetikzlibrary[patterns.meta] % ConTEXt when using TikZ
\usepackage[acronym]{glossaries}
\newacronym{rl}{RL}{reinforcement learning}
\newacronym{fl}{FL}{federated learning}
\newacronym{dl}{DL}{distributed learning}
\newacronym{pqos}{PQoS}{Predictive Quality of Service}
\newacronym{eu}{EU}{European Union}
\newacronym{lidar}{LiDAR}{Light Detection and Ranging}
\newacronym{v2x}{V2X}{Vehicle-to-Everything}
\newacronym{ran}{RAN}{Radio Access Network}
\newacronym{gnb}{gNB}{Next Generation NodeB}
\newacronym{ml}{ML}{machine learning}
\newacronym{ai}{AI}{artificial intelligence}
\newacronym{ns3}{ns-3}{Network Simulator}
\newacronym{qos}{QoS}{Quality of Service}
\newacronym{qoe}{QoE}{Quality of Experience}
\newacronym{cr}{C-R}{Compression Raw}
\newacronym{csc}{C-SC}{Compression Segmentation Conservative}
\newacronym{csa}{C-SA}{Compression Segmentation Aggressive}
\newacronym{phy}{PHY}{Physical Layer}
\newacronym{mac}{MAC}{Medium Access Control}
\newacronym{rlc}{RLC}{Radio Link Control}
\newacronym{pdcp}{PDCP}{Packet Data Convergence Protocol}
\newacronym{sinr}{SINR}{Signal to Interference plus Noise Ratio}
\newacronym{mdp}{MDP}{Markov Decision Process}
\newacronym{dnn}{DNN}{Deep Neural Network}
\newacronym{dqn}{DQN}{Deep Q Learning}
\newacronym{ddqn}{D-DQN}{Double Deep Q-Learning Network}
\newacronym{app}{APP}{Application}
\newacronym{kpi}{KPIs}{Key Performance Indicators}
\newacronym{cd}{CD}{Chamfer Distance}
\newacronym{prr}{PRR}{Packet Reception Rate}
\newacronym{mcs}{MCS}{Modulation Coding Scheme}
\newacronym{ofdm}{OFDM}{Orthogonal Frequency-Division Multiplexing}
\newacronym{imsi}{IMSI}{International Mobile Subscriber Identity}
\newacronym{pdu}{PDUs}{Packet Data Unit}
\newacronym{sgd}{SGD}{Stochastic Gradient Descent}
\newacronym{3gpp}{3GPP}{3rd Generation Partnership Project}
\renewcommand{\figurename}{Fig.}
\usepackage[font=scriptsize]{subcaption}
\usepackage[font=footnotesize]{caption}

\begin{document}
%+++++++++++++++++++++++++++++++++++++++++++
\title{Towards Decentralized Predictive Quality of Service in Next-Generation Vehicular Networks}
\author{
\IEEEauthorblockN{Filippo Bragato, Tommaso Lotta, Gianmaria Ventura   \\ Matteo Drago, Federico Mason, Marco Giordani, Michele Zorzi} 
\IEEEauthorblockA{Department of Information Engineering, University of Padova, Italy.}
%\\Email: \texttt{\{name.surname\}@dei.unipd.it}}\\
\thanks{Matteo Drago is now a Research Engineer at Delart Technology Services LLC, 312 Arizona Ave., Santa Monica, CA 90401. Federico Mason is now a Postdoctoral Researcher at Department of Biomedical and Neuromotor Sciences, University of Bologna, Italy. }}

%+++++++++++++++++++++++++++++++++++++++++++

\maketitle

%\IEEEaftertitletext{\vspace{-3ex}}

\begin{abstract}
To ensure safety in teleoperated driving scenarios, 
communication between vehicles and remote drivers must satisfy strict latency and reliability  requirements.
In this context, \gls{pqos} was investigated as a tool to predict unanticipated degradation of the \gls{qos}, and allow the network to react accordingly. %in order to guarantee a target performance. 
%In previous studies, \gls{rl} has been identified as a possible means to perform accurate predictions.
In this work, we design a \mbox{\gls{rl}} agent to implement \gls{pqos} in vehicular networks. To do so, based on data gathered at the \gls{ran} and/or the end vehicles, as well as \gls{qos} predictions, our framework is able to identify the optimal level of compression to send automotive data under low latency and reliability constraints.
%To do so, based on data gathered from the \gls{ran} and/or end vehicles, and identifies the optimal level of compression to send automotive data 
We consider different learning schemes, including centralized, fully-distributed, and federated learning. We demonstrate via ns-3 simulations that, while centralized learning generally outperforms any other solution, decentralized learning, and especially federated learning, offers a good trade-off between convergence time and reliability, with positive implications in terms of privacy and complexity. %needed for the \gls{rl} agent to converge.
\end{abstract}

\begin{tikzpicture}[remember picture,overlay]
\node[anchor=north,yshift=-20pt] at (current page.north) {\parbox{\dimexpr\textwidth-\fboxsep-\fboxrule\relax}{
\centering\footnotesize This paper has been accepted for publication at IEEE Information Theory and Applications Workshop (ITA). 2023 {\textcopyright}IEEE.\\     
Please cite it as: F. Bragato, T. Lotta, G. Ventura M. Drago, F. Mason, M. Giordani, M. Zorzi, ``Towards Decentralized Predictive Quality of Service in Next-Generation Vehicular Networks,'' \emph{IEEE Information Theory and Applications Workshop (ITA)}, 2023.}};
\end{tikzpicture}

\glsresetall
\section{Introduction} \label{sec:introduction}
\begin{comment}
\begin{figure*}[th!]
\centering
    \begin{subfigure}{0.2\textwidth}
        \centering
        \includegraphics[width=\textwidth]{images/centralized}
        \caption{Centralized}
        \label{fig:centr}
    \end{subfigure}\qquad \qquad \qquad
    \begin{subfigure}{0.2\textwidth}
        \centering
        \includegraphics[width=\textwidth]{images/distributed}
        \caption{Distributed}
        \label{fig:distr}
    \end{subfigure}\qquad \qquad \qquad
    \begin{subfigure}{0.2\textwidth}
        \centering
        \includegraphics[width=\textwidth]{images/federated}
        \caption{Federated}
        \label{fig:fed}
    \end{subfigure}
    \caption{Representation of different approaches.}
    \label{fig:schemes}
\end{figure*}
\end{comment}

\begin{figure*}[th!]
\centering
\setlength\belowcaptionskip{-0.33cm}
        \includegraphics[width=0.95\textwidth]{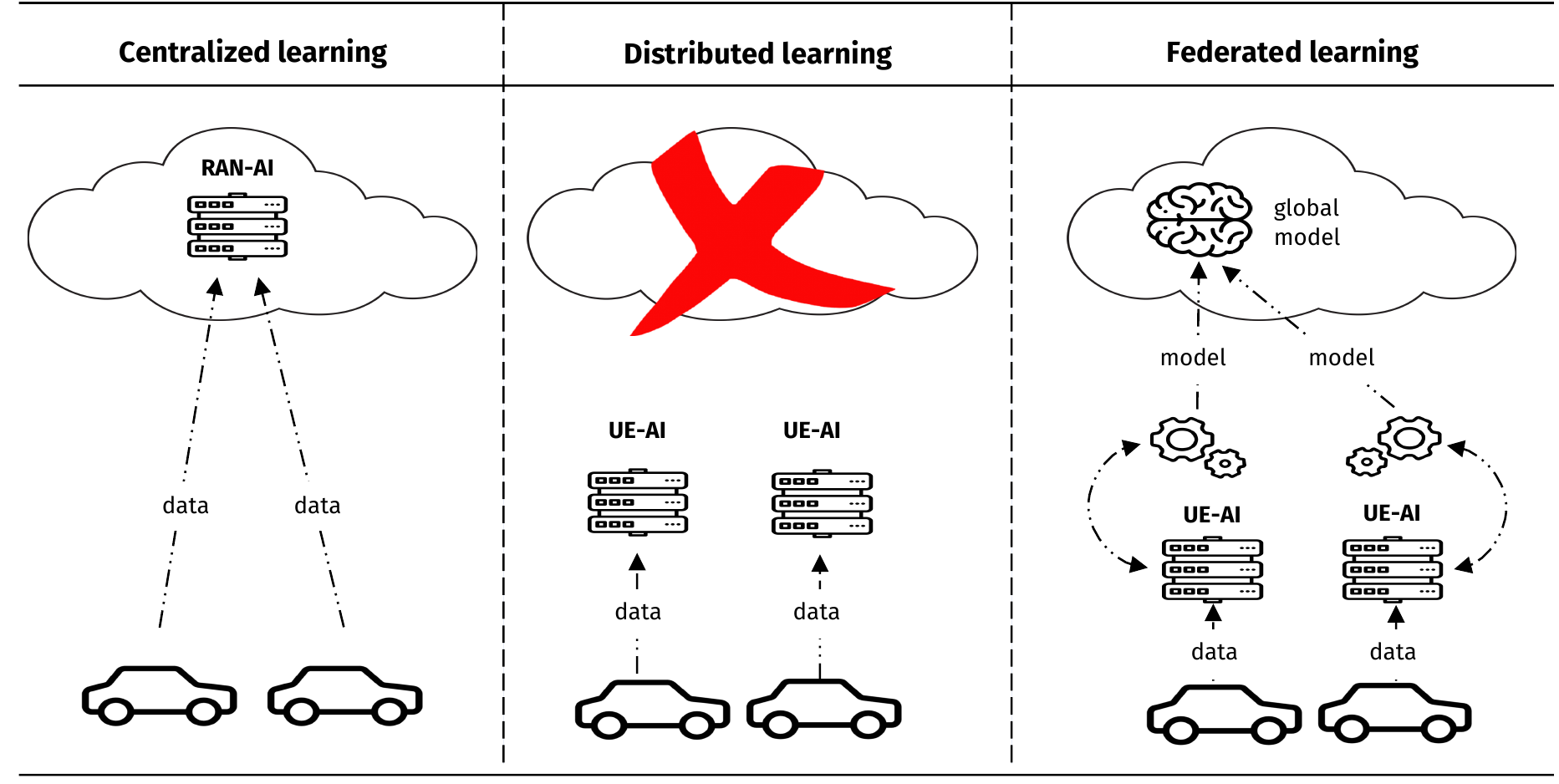}
        \caption{Representation of different PQoS architectures based on reinforcement learning.  Centralized learning (left) uses the ``RAN-AI'' entity at the gNB~\cite{mason2022reinforcement,drago2022artificial} to collect (and train on) global data from all the vehicles in the network; distributed learning (center) uses the ``UE-AI'' entity to collect (and train on) local data; federated learning (right) uses the ``UE-AI'' entity to collect (and train on) local data, as well as on a global model trained at the gNB.}
        \label{fig:scheme}
    \end{figure*}

Over the past few years, autonomous driving has been studied as a means to improve road safety, as research indicates that more than 75\% of  road crashes are due to human error~\cite{STANTON2009227}. 
% As such, the introduction of autonomous driving is seen as a means to drastically limit them. 
Furthermore, intelligent vehicles could help limit fuel consumption and carbon emissions via efficient driving optimizations performed by \gls{ai} agents~\cite{9097944}. To acknowledge this trend, the \gls{eu} has started the regulation of autonomous driving~\cite{eu-2144-2019},  thus promoting faster developments in this~domain. 

% Recent advancements in unsupervised learning models, object detection techniques and sensors manufacturing, have further fueled research in this field.
However, despite recent advances in the automotive industry, truly autonomous driving without human intervention is still far from reality due to fundamental technical challenges~\cite{zhang2020toward}.
On the other hand, the research community is exploring teleoperated driving, which enables the control of vehicles by a remote driver that can be either human or software.
%Teleoperated driving could completely change the way of working in zones with limited access areas such as ports and big factories or dangerous areas such as building sites and mines~\cite{LEE2022104119}.
%A successful and efficient deployment of this new paradigm could completely change the way work is carried out in zones with limited access areas (e.g., ports, big factories) or dangerous zones, as remote building sites and mines~\cite{LEE2022104119}. In these cases, operators could easily drive even multiple vehicles at once without moving from their offices, increasing workers' safety and improving the logistic.
In order to enable teleoperated driving, the remote driver needs to receive different measurements from the vehicle, including perception data of the environment, which are generally obtained through onboard sensors like high-resolution cameras, depth cameras, and \gls{lidar} sensors.
\glspl{lidar}, in particular, generate a 3D representation of the environment in the form of a point cloud, and can be used for detection and recognition of road entities~\cite{rossi2021role}. %nardo2022point.
% For example \gls{lidar} sensors are equipped in vehicles such as the Honda Legend and the Mercedes-Benz S-Class which are the first to have reached level 3 automation in the market.
However, \gls{lidar} data may cause network congestion and delays. For example, for raw uncompressed high-resolution LIDAR perceptions, the average file size is around 25 Mbits which, at 10 perceptions/s, produce an average data rate of around 250 Mbps~\cite{nardo2022point}.
Similar data rates may be challenging to handle for resource-constrained networks, considering that teleoperated driving comes with tight  requirements, especially low latency to increase the responsiveness of the driver and high reliability to receive accurate driving~commands.

In this context, \gls{pqos} can provide \gls{v2x} systems with advance notifications in case network requirements are not satisfied, and allow the network to react accordingly~\cite{boban2021predictive}.
Specifically, \gls{pqos} can foresee unanticipated \gls{qos} degradation (due to, for example, scarce coverage or congested networks), and guarantee more reliable~driving.

Recently, it was demonstrated that \gls{ml} is a valid tool to predict and optimize wireless networks~\cite{wang2017machine}, unlike deterministic methods such as time-series analytics or statistical methods.
Notably, a regression problem can be formulated to predict when future failures will occur, while classification or clustering problems can predict what kind of failure is more likely to take place in certain network conditions~\cite{boutaba2018comprehensive}.
Among other techniques, \gls{rl} has been successfully implemented to support \gls{pqos} in \gls{v2x}. For example, in our previous work~\cite{mason2022reinforcement} we developed a centralized framework based on a new network entity called ``RAN-AI'' (Fig.~\ref{fig:scheme} left) that, connected to the \gls{ran}, predicts the behavior of the network and optimizes driving decisions accordingly~\cite{drago2022artificial}. 

% \begin{figure}[b]
% \centering
%     \begin{subfigure}{0.3\textwidth}
%         \centering
%         \includegraphics[width=\textwidth]{images/centralized2.png}
%         \caption{Centralized}
%         \label{fig:centr}
%     \end{subfigure}
%     \begin{subfigure}{0.3\textwidth}
%         \centering
%         \includegraphics[width=\textwidth]{images/distributed2.png}
%         \caption{Distributed}
%         \label{fig:distr}
%     \end{subfigure}
%     \begin{subfigure}{0.3\textwidth}
%         \centering
%         \includegraphics[width=\textwidth]{images/federated2.png}
%         \caption{Federated}
%         \label{fig:fed}
%     \end{subfigure}
%     \caption{Representation of different approaches.}
%     \label{fig:schemes}
% \end{figure}

However, centralization requires continuous exchange of data (driving commands) to (from) the \gls{ran} at the \gls{gnb}, which may not be compatible with low latency in \gls{v2x} scenarios~\cite{pase2022distributed}. 
Therefore, in this work we explore the feasibility of  decentralized/distributed PQoS. Specifically, we extend the RAN-AI framework, and implement a new entity called ``UE-\gls{ai}'' that operates directly at the user/vehicle level, rather than at the RAN, enabling \gls{dl} and \gls{fl}~\cite{9141214}. Notably, the UE-AI integrates an \gls{rl} agent to choose the optimal level of compression to send LiDAR data and
satisfy \gls{qos} requirements.
While in \gls{dl} (Fig.~\ref{fig:scheme} center) each vehicle trains its own \gls{rl} agent using only onboard data,   \gls{fl} (Fig.~\ref{fig:scheme} right) still keeps data localized, but requires vehicles to send the results of the training process to the \gls{gnb}, where results are combined in a new global model that vehicles can use to improve their local agents iteratively.
%each vehicle still trains its own \gls{rl} model using local data, but then sends the results of the training to the \gls{gnb}, which combines those results in a new model and forwards it back to vehicles. Each vehicle then replaces its model with the newest one and continues the process.
The UE-\gls{ai} prevents end users from exchanging network data with the \gls{gnb}, except at most the trained model, thus reducing the communication overhead. % between the end users and the \gls{gnb}, thus increasing reliability and reducing the latency.
Among other things, this approach promotes privacy, since no user-sensitive data are disseminated through the network, which is another important aspect to consider in the design of safety-critical V2X networks.

The performance of our UE-AI design for decentralized PQoS is evaluated in ns-3 against a centralized benchmark.
The results demonstrate that, while centralized PQoS generally outperforms any other solution, decentralized PQoS can still improve the QoS of V2X applications compared to other baselines that do not implement RL techniques.
%a good trade-off between convergence time and reliability.
Unlike distributed learning, federated learning works well even in congested networks, while still promoting~privacy.

%Under federated learning, multiple people remotely share their data to collaboratively train a single deep learning model, improving on it iteratively, like a team presentation or report. Each party downloads the model from a datacenter in the cloud, usually a pre-trained foundation model. They train it on their private data, then summarize and encrypt the model’s new configuration. The model updates are sent back to the cloud, decrypted, averaged, and integrated into the centralized model. Iteration after iteration, the collaborative training continues until the model is fully trained.

The paper is organized as follows. Sec.~\ref{sec:ns3tech} presents our UE-AI implementation, Sec.~\ref{sec:ReinforcementLearningModel} describes our \gls{rl} models, Sec.~\ref{sec:performance} provides numerical results, while Sec.~\ref{sec:concl} concludes the paper with suggestions for future~research.

%\newpage

\section{Implementation of Decentralized PQoS} \label{sec:ns3tech}

This work extends the \gls{ran}-\gls{ai} framework presented in~\cite{mason2022reinforcement,drago2022artificial} where PQoS functionalities are centralized at the RAN, to propose a new UE-\gls{ai} entity implementing distributed and federated \gls{pqos}, as described below.

\subsection{Application Model}
\label{sub:app}

We consider a teleoperated driving scenario where connected vehicles are controlled by a remote driver.
Specifically, vehicles acquire and disseminate \gls{lidar} perceptions of the surrounding environment, to which drivers reply with ad hoc driving instructions.

As described above, sending \gls{lidar} data through the network may be challenging as it can easily congest the wireless channel, especially in case of poor communication performance. 
In this case, \gls{lidar} data can be compressed before transmission to reduce the file size and alleviate network congestion~\cite{nardo2022point}.
To this aim, PQoS allows vehicles to choose the optimal level of compression to  satisfy communication requirements in terms of latency and reliability, based on QoS predictions in future time instants and locations.

%in order to satisfy communication requirements in terms of latency and reliability, vehicles may change the level of compression of  data before transmission, based to the output of the \gls{qos} inference In fact, increasing the compression level of a \gls{lidar} point cloud would lead to a smaller packet size and, eventually, to a decrease in network congestion and higher \gls{qos}.

Our reference pipeline for data compression is based on the Hybrid Semantic Compression (HSC) algorithm proposed in~\cite{9500523}. First, data are segmented using  RangeNet++~\cite{milioto2019rangenet++} to identify critical objects in the scene. Then, data are compressed using Draco~\cite{draco}, choosing among 5 levels of compression and 14 levels of quantization. 
In this work, after a thorough preliminary study, we limit the algorithm to three alternatives:
\begin{itemize}
    \item {\gls{cr}}: segmentation is not applied, and compression is done on the original point~cloud. % The average file size is 16 times lower than.
    \item {\gls{csc}}: the points associated to the road elements are removed from the point cloud after segmentation, before compression. %On average, \gls{scs} reduces the file size compared to raw LiDAR data by 30 times.
    \item {\gls{csa}}: the points associated to buildings, vegetation, and the background are also removed from the point cloud, before compression. The resulting point cloud consists only of dynamic elements such as pedestrians and vehicles. %On average, \gls{scs} reduces the file size compared to raw LiDAR data by 180 times.
\end{itemize}
%It has to be highlighted that this processing is carried out offline, and the input of the simulation consists of trace files of packet sizes associated to different environment scenes with different compression options. 

In ns-3, HSC is implemented in the \texttt{KittiTraceBurstGenerator} application, as described in~\cite{drago2022artificial}, which uses real LiDAR data from the Kitti dataset~\cite{geiger2012are}, and applies compression and/or segmentation when appropriate. Then, the \texttt{BurstyApplication} and  \texttt{BurstSink} classes are used to fragment, transmit, receive, and aggregate packets.

\subsection{The UE-AI Entity}
\label{sub:ue-ai}

%In the following we assume that end users have access to data available in the \gls{ran} through a dedicated feedback channel; a possible future improvement of this work would include a real implementation of this feedback, so that the data availability would also be impacted by the network congestion and the channel quality. 

%The UE-AI implements the following functionalities to support decentralized PQoS:
%\begin{itemize}
%\item Collect network metrics at different layers. % specific to the single user/vehicle where the UE-AI is installed.
%\item Share network metrics with the local \gls{rl} agent (either distributed or federated), which (i) makes QoS predictions, and (ii) returns the optimal compression level to satisfy QoS requirements, relative to the user/vehicle where the UE-AI is installed. The \gls{rl} model will be described in Sec.~\ref{sec:ReinforcementLearningModel}.
%\item Change the compression level at the application, based on the decisi on of the RL agent.
%\end{itemize}

In ns-3, the UE-AI entity is fully integrated with the \texttt{mmwave} module~\cite{mezzavilla2018end}, and is implemented in the new \texttt{UeAI} and \texttt{MmWaveUeNetDevice} classes to support decentralized PQoS. Notably, it consists of the following~methods.

\paragraph{Initialization}
The \texttt{InstallUserAI} method is used to install the UE-AI on a specific user/vehicle, initialize measurements collection, and schedule status updates.

\paragraph{Measurement collection}
The \texttt{RxPacket\-TraceUe} method collects metrics at the Physical (PHY) layer, specifically:
    \begin{itemize}
        \item \gls{mcs} of the transmission;
        \item Number of transmitted \gls{ofdm} symbols;
        \item Avg. \gls{sinr}. 
    \end{itemize}
    
Then, the {\texttt{SendStatusUpdate}} method collects metrics at the \gls{rlc}, \gls{pdcp}, and \gls{app} layers, specifically:
    \begin{itemize}
        \item \gls{imsi}, which uniquely identifies each  user/vehicle;
%\item Transmitted and received bytes; 
\item Minimum, maximum, mean, and standard deviation of the \gls{pdu} delay and size;
\item Number of transmitted and received \gls{pdu};
%\item Min. and max. experienced delay;
\item Minimum and maximum size of the received packets.
%\item Transmitted and received packet bursts;
%\item Transmitted and received bytes;
%\item Mean and std. deviation of packet burst delay;
%\item Min. and max. experienced latency.
    \end{itemize}
%\paragraph{\texttt{MmWaveUeNetDevice::InstallUserAI}} It installs the UE-AI entity on a specific user/vehicle, initializes measurements collection, and schedules  status updates.

Compared to the RAN-AI entity in~\cite{mason2022reinforcement,drago2022artificial}, which gathers measurements from all users/vehicles connected to the network, the UE-AI entity is executed per-user, and measurements are referred to the single user/vehicle where the UE-AI is installed.
While this approach permits the RAN-AI to achieve better perception of the environment, and thus take more accurate optimizations, it requires coordination with the end users via dedicated control signaling, which may involve additional delays and complexity.

\paragraph{Network control}
The \texttt{SendStatusUpdate} method calls the \texttt{ReportMeasures} method, which forwards network measurements to the \gls{rl} agent (distributed or federated) which (i) 
%and receives instructions accordingly. Specifically, the \gls{rl} agent 
makes QoS predictions, and (ii) obtains the optimal compression level for LiDAR data to satisfy QoS requirements, relative to the user/vehicle where~the UE-AI is installed. 

In ns-3, RL functionalities are integrated in the \texttt{UeAI} class via the \texttt{ns3-ai} extension~\cite{hao2020ns3ai}, which provides efficient and high-speed data exchange between AI/RL algorithms (implemented in Python) and ns-3 (implemented in C++). 
The RL model will be described in Sec.~\ref{sec:ReinforcementLearningModel}.
%\item Change the compression level at the application, based on the decision of the RL agent.

\paragraph{Application control}
The \texttt{NotifyActionIdeal} method propagates the RL agent's decision (i.e., the optimal action) to the application to be configured~accordingly.
%where the UE-AI changes the compression level at the application based on the decision of the RL agent.

%propagates the received instructions to the application.

\section{Reinforcement Learning Models} \label{sec:ReinforcementLearningModel}
As introduced in Sec.~\ref{sec:ns3tech}, the UE-AI entity implements an \gls{rl} agent that, in the attempt to satisfy QoS requirements, obtains the optimal compression level to apply to LiDAR data before transmission. 
Notably, the \gls{rl} model is a powerful mathematical framework that characterizes the target environment as a \gls{mdp}. 
In our scenario, time is discretized in steps $t=0,1,2,...$ and, at each step $t$, the agent observes the state of the environment $s_t \in \mathcal{S}$, and chooses an action $a_t \in \mathcal{A}$, where $\mathcal{S}$ and $\mathcal{A}$ are the state and action spaces, respectively.
Depending on $s_t$ and $a_t$, the environment evolves into a new state $s_{t+1} \in \mathcal{S}$, while the agent receives a reward $r_t \in \mathbb{R}$ accordingly.

In an \gls{rl} scenario, the goal is to find the optimal policy $\pi^*:\mathcal{S} \rightarrow \mathcal{A}$ to maximize the cumulative sum of the rewards $\sum_{\tau=t}^{+\infty} \gamma^{\tau-t} r_{\tau}$ obtained over time~\cite{kaelbling1996reinforcement}, where $\gamma \in [0, 1)$ is the discount factor.
%In particular, the reward is discounted by a factor $\gamma \in [0, 1]$ that makes current decisions more important with respect to future ones.
In Sec.~\ref{sub:learning-alg} we formalize our RL algorithm, in Sec.~\ref{sec:reward} we describe the reward function, and in Sec.~\ref{subsec:approaches} we present several RL schemes for PQoS.

\subsection{Learning Algorithm}
\label{sub:learning-alg}

In this work, the optimal policy is learned through the \gls{ddqn} algorithm~\cite{vanHasselt_Guez_Silver_2016}, which approximates the agent's policy $\pi$ by a \gls{dnn} called primary network. 
At each step $t$, the primary network receives the current state $s_t$ as input, and returns the Q-value $Q_{\pi}(a_t,s_t)$ for each possible action $a_t \in \mathcal{A}$.
The value of $Q_{\pi}(a_t, s_t)$ represents the expected cumulative reward that the agent will receive by taking action $a_t$ in the current state $s_t$, and following policy $\pi$.

According to the \gls{ddqn} algorithm, we implement an additional \gls{dnn} called {target network}. 
The target network is an asynchronous version of the primary network, whose parameters are updated at regular intervals, i.e., learning steps, $\upsilon$.
During the agent's training phase, the algorithm exploits the two \glspl{dnn} to estimate the current and future Q-values \cite{vanHasselt_Guez_Silver_2016}.
This approach avoids the overestimation of Q-values, ensuring a more robust learning process.

In order to improve the system convergence, we introduce a {memory replay} approach.
Hence, we provide the agent with an internal memory of up to $\mu$ transitions $(s_t,a_t,r_t,s_{t+1})$.
At each step $t$, the agent randomly picks $\beta$ transitions from the memory, which are used to update the primary network accordingly.
When the memory capacity is reached, the oldest transitions are discarded, so that the learning phase is based on the most recent observations.

Finally, to encourage the exploration of the state and action spaces, we implement an $\epsilon$-greedy policy.
Therefore, at each step $t$, the agent chooses a random action $a_t \in \mathcal{A}$  with probability $\epsilon \in [0,1]$ (exploration), while with probability $1-\epsilon$ the agent chooses the best current action (exploitation); the value of $\epsilon$ decreases linearly during the training.

\subsection{Reward Function} \label{sec:reward}
The RL model is characterized by two main components: the state and the reward function.
In this work, the state~$s$ includes all the measurements collected by the UE-AI, including metrics at the PHY, RLC, PDCP, and APP layers of the local user/vehicle.
The reward $r$ depends on both:
%\gls{qos} and \gls{qoe} considerations:
\begin{itemize}
    \item \acrfull{qos}, i.e., the end-to-end communication delay $\delta_{\text{APP}, t}$  for a vehicle at time $t$ should be lower than the maximum tolerable delay $\delta_{\text{M}}$.
    \item \gls{qoe}, i.e., the transmitted data should be accurate enough to perform driving operations, e.g., object detection. In this work, the \gls{qoe} is inversely proportional to the symmetric point-to-point \gls{cd}~\cite{9500523} between the original data $\mathcal{P}$ acquired by the \gls{lidar} and the transmitted data $\hat{\mathcal{P}}$ after compression and/or segmentation, where:
\begin{equation}
\text{CD} = \sum_{\forall \textbf{p} \in \mathcal{P}} \min_{\forall \hat{\textbf{p}} \in \hat{\mathcal{P}}}  \left\Vert \textbf{p} - \hat{\textbf{p}} \right\Vert_2^2 + \sum_{\forall \hat{\textbf{p}} \in \hat{\mathcal{P}}} \min_{\forall \textbf{p} \in \mathcal{P}}  \left\Vert \textbf{p} - \hat{\textbf{p}} \right\Vert_2^2.
\end{equation}
\end{itemize}
Practically, the reward $r_t$ received at step $t$ is given by
%a weighted sum of two parameters, i.e., $\sigma_{t}^{s}$ and $\sigma_{t}^{e}$.
\begin{equation}
\label{eq:reward}
    r(\sigma_{t}^{s}, \sigma_{t}^{e}) =
    \begin{cases}
    -1  & \text{if } \sigma_{t}^{s}>1; \\
     1 - 2 \alpha \sigma_{t}^{e} - 2 (1 - \alpha) \sigma_{t}^{s} & \text{otherwise,}
    \end{cases}
\end{equation}
where $\alpha \in [0,1]$ is a tuning parameter which regulates the trade-off between the QoS ($\sigma_{t}^{s}$) and QoE ($\sigma_{t}^{e}$) components. 

In Eq.~\eqref{eq:reward}, $\sigma_{t}^{s}$ and $\sigma_{t}^{e}$ both take values in $[0,1]$, so that $r(\cdot):  [0,1] \times  [0,1] \rightarrow [-1, 1]$.
In particular, $\sigma_{t}^{s}$ is given by
\begin{equation}
    \sigma_{t}^{s} = {\delta_{\text{APP}, t}}/{\delta_{\text{M}}},
\end{equation}
while $\sigma_{t}^{e}$ depends on the maximum tolerable \gls{cd} (CD$_\text{M}$) and a simulation parameter called \textit{penalty} ($\rho$), and is given~by
\begin{equation}
\label{eq:qoe_value}
    \sigma_{t}^{e} = {\text{CD}_{t}}/({\text{CD}_{\text{M}} + \rho}).
\end{equation}

\subsection{Learning Schemes}
\label{subsec:approaches}
We investigate three alternatives to perform \gls{pqos}: centralized (our benchmark based on the implementation in~\cite{mason2022reinforcement,drago2022artificial}), distributed, and federated.

\paragraph{Centralized PQoS}

The RAN-AI is used to train a single global agent at the \gls{gnb}, exploiting measurements collected by all the users/vehicles in the network, which improves the convergence time of the learning process. 

On the downside, centralization requires all the users/vehicles to share local data with the gNB, as well as the gNB to send the agent's decisions to the users/vehicles, which may cause congestion and delay in the network. 
At the same time, data may be exposed to security risks at the gNB, including accidental or illegitimate destruction, loss, alteration, or unauthorized access to sensitive data, that could be used, for example, to track the users' positions or falsify driving commands.

\paragraph{Distributed PQoS}

Each vehicle implements an independent agent via the UE-AI, which is trained only with onboard data gathered locally.
On one side, data is kept local, thus ensuring privacy and reducing communication overhead and delays to/from the gNB during the learning process.
On the other side, the algorithm convergence is slower: hence, at each simulation step, the UE-AI provides the agent with data from a single user/vehicle, regardless of how many vehicles are deployed, which in turn needs more interactions with the system to optimize its decisions. 

\paragraph{Federated PQoS}
%Under federated learning, multiple people remotely share their data to collaboratively train a single deep learning model, improving on it iteratively, like a team presentation or report. Each party downloads the model from a datacenter in the cloud, usually a pre-trained foundation model. They train it on their private data, then summarize and encrypt the model’s new configuration. The model updates are sent back to the cloud, decrypted, averaged, and integrated into the centralized model. Iteration after iteration, the collaborative training continues until the model is fully trained.
Each vehicle still implements an independent agent via the UE-AI, which is trained on local data. In addition, vehicles periodically share intermediate learning model updates to a central server (e.g., located at the gNB), which trains a global model through the weighted average of the received data, improving on it iteratively~\cite{9084352}.
In particular, the weights of the average are proportional to the actual number of learning steps performed by each local agent.
%The UE-AI trains learning models over remote users/vehicles that keep raw data localized,  with only intermediate updates being sent with a central server, which may be located at the gNB~\cite{9084352}.
%While many works in the literature have investigated \gls{fl} solutions for supervised learning, the field of federated \gls{rl} has not yet been fully explored yet~\cite{8791693}.
%each vehicle implements an independent agent trained according to local information while a central system periodically triggers a synchronization of the local agents.
%every $\upsilon$ steps, each vehicle shares the weights of its local model with the \gls{gnb}, which updates a global model through a weighted average of the received information.
Then, every $\upsilon$ steps, each vehicle downloads the pre-trained global model from the gNB. They train it on their local data, then summarize and encrypt the model's new configuration. 
The model updates are sent back to the gNB, decrypted, averaged, and integrated into the global model. Iteration after iteration, the collaborative training continues until the model is fully trained.

%The global model is transmitted to every single vehicle, which replaces its local model with that computed by the \gls{gnb}. 
The main advantage of federated learning is that only model updates are communicated to the gNB, while raw data are kept locally, thus promoting privacy.
Also, the availability of the global model at the gNB implies more data for the agents to optimize their decisions, which translates in faster convergence compared to distributed~PQoS~\cite{pase2021convergence}.
%Moreover, the described framework enables a strong reduction of the communication overhead between the vehicles and the \gls{gnb}, possibly increasing the overall performance.

Notice that, while many works in the literature have investigated federated learning solutions for supervised learning, the field of federated \gls{rl} has not yet been fully explored~\cite{8791693}, which motivates our analysis.

\section{Performance Evaluation}
\label{sec:performance}

In this section, after introducing our system parameters (Sec.~\ref{sub:params}), we validate the performance of the PQoS schemes in Sec.~\ref{subsec:approaches} via ns-3 simulations (Sec.~\ref{sub:results}). 
%we implement the different learning approaches presented in Sec.~\ref{sec:ReinforcementLearningModel} in the \gls{pqos} framework described in~\ref{sec:ns3tech}, considering a network scenario with multiple vehicles sharing the same communication channel.  Our goal is to provide the system users with an efficient communication policy that can address the trade-off between \gls{qos} and \gls{qoe}. In particular, the \gls{qoe} is measured as a function of the \gls{cd}, while the \gls{qos} is related to \gls{v2x}-specific \gls{kpi} for teleoperated driving.

\subsection{Simulation Parameters}
\label{sub:params}
Simulation parameters are in Tab.~\ref{tab:params}, and described below.

\paragraph{Scenario}
We consider a \gls{v2x} scenario where $n$ vehicles are deployed and operate at a carrier frequency of $3.5$~GHz and with a total bandwidth of $50$~MHz, while the transmit power is set to $23$~dBm.

\paragraph{Application}
Based on the \gls{3gpp} specifications for teleoperated driving, we set the maximum tolerable delay to $\delta_{\rm M}=50$~ms and the maximum  tolerable CD to CD$_{\rm M}=45$.
At the application, each vehicle generates LiDAR data at a rate of 10 perceptions/s, while the average perception size depends on the compression level and is 200 KB for C-R, 104 KB for C-SC, and 17 KB for C-SA~\cite{9500523}.

\begin{table}[t!]
\centering
\scriptsize
\renewcommand{\arraystretch}{1.3}
\caption{Scenario parameters.}
\label{tab:params}
\begin{tabular}{l|l|l}
  \toprule
  Parameter & Description & Value \\
  \hline
  
  $f_c$ & Carrier frequency & 3.5 GHz       \\
  $B$ & Total bandwidth                           & 50 MHz        \\
  $P_{\rm TX}$ & Transmission power & 23 dBm        \\ 
  $n$ & Number of vehicles & $\{1,\,5,\,8\}$ \\ \hline
  $\delta_{\rm M}$ & Max. tolerated delay    & $50$ ms     \\ 
  $\text{CD}_{\text{M}}$ & Max. tolerated Chamfer Distance  & 45     \\
  $f_{\rm C-R}$ & {LiDAR file size C-R} & 200 KB \\
  $f_{\rm C-SC}$ & {LiDAR file size C-SC} & 104 KB \\
  $f_{\rm C-SA}$ & {LiDAR file size C-SA} & 17 KB \\
  $r$ & {LiDAR perception rate} & 10/s \\
  \hline
   $N_{\text{h}}$ & Hidden dimensions & $16\times64$ \\
        $\gamma$ & Discount factor & 0.95 \\
        $\upsilon$ & Update interval & $0.1$ s \\
        $\mu$ & Memory replay size & $8 \cdot 10^4$ B \\
        $\beta$ & Batch size & 32 B\\
        $\zeta$ & Learning rate & $10^{-5}$\\
%  \multicolumn{2}{c|}{Layer size (inputs $\times$ outputs) -- ReLU activation} & $12\times 6$ \\ \hline
  %\multicolumn{2}{c|}{Layer size (inputs $\times$ outputs) -- Linear activation} & $6\times 3$ \\ 
  \bottomrule
  \end{tabular}
\end{table}

\paragraph{Learning model}
During the training phase, we consider $80$ steps per episode (where an episode is equivalent to an independent simulation), and we set the number of total episodes to $3\,000$. 
Every step has a fixed duration of $100$~ms, leading to a total episode duration of $8$~s. 
During the testing phase, instead, we consider a total of $100$ episodes with $800$ steps each. 
Finally, the \gls{ddqn} algorithm is approximated by a \gls{dnn} architecture with two hidden layers with $16$ and $64$ neurons respectively. The \gls{dnn} uses the \gls{sgd} algorithm with $\zeta=10^{-5}$ as the learning rate.
The other parameters relative to the system training are given in Tab.~\ref{tab:params}.

\paragraph{Simulations}
The performance evaluation is done using ns-3, a popular discrete-event simulator for networks~\cite{henderson2008network}. Compared to other competitors, ns-3 incorporates accurate models of the whole 5G NR protocol stack, and enables scalable end-to-end simulations. 
We consider centralized, distributed, and federated learning for PQoS, versus a baseline in which the compression level (either C-R, S-SC, S-SA) is set a priori and does not change during the simulation. Performance results are given as a function of the number of users, and in terms of the reward gained by the agent and the average QoS and QoE of the users/vehicles. 

\begin{figure}[t!]
    \centering
    \includegraphics[width=0.95\columnwidth]{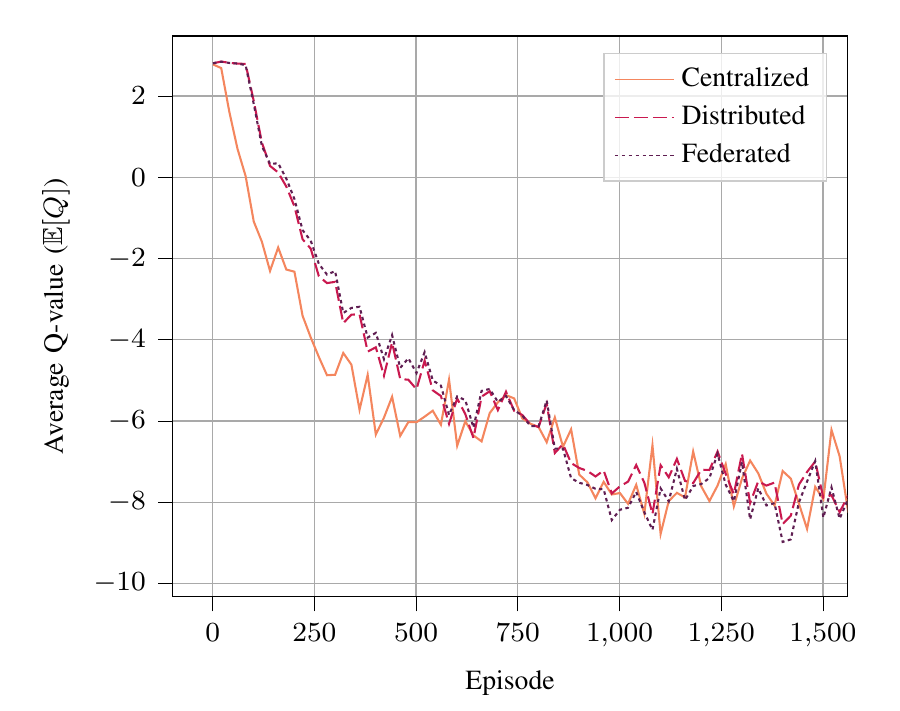}
    \caption{Average Q-value during training for different PQoS schemes (centralized, distributed, federated), and $n=3$.}
    \label{fig:q-values-approach}
\end{figure}

\begin{figure}[t!]
    \centering
    \setlength\belowcaptionskip{-0.5cm}
    \includegraphics[width=0.95\columnwidth]{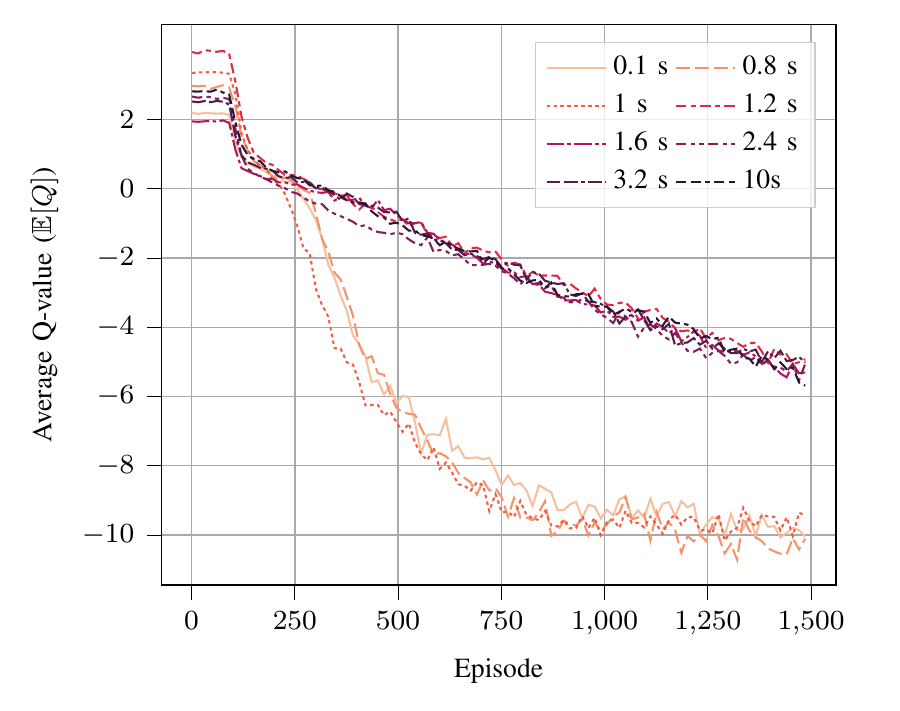}
    \caption{Average Q-value during training for federated PQoS, for different values of the learning step $\upsilon$. We set $n=5$.}
    \label{fig:q-values}
\end{figure}

\subsection{Numerical results}
\label{sub:results}
At first, we study the convergence time of the three PQoS schemes (centralized, distributed, federated) described in Sec.~\ref{subsec:approaches}.
To do so, in Fig.~\ref{fig:q-values-approach} we represent the average Q-value for each scheme, considering a scenario with $n=3$ vehicles and a training phase of $1\,500$ episodes. 
During the first $500$ episodes, centralized PQoS is the only scheme to converge in our experiments. 
In fact, centralized PQoS can leverage data from all the users/vehicles in the network during the training phase (vs. local data for decentralized PQoS), and can optimize faster.
Indeed, this approach enables a better exploration of the state and action spaces, making it a more convenient solution to deploy in case of a limited training time. 
On the other hand, after a sufficient number of episodes,
the three PQoS schemes present similar performance, and reach full convergence at the same time, i.e., after around $1\,000$ episodes. 
%As such, decentralized learning stands out as an interesting approach to 

We recall that the performance of federated PQoS depends on the periodicity $\upsilon$ at which training model updates are sent to the gNB for processing. To investigate this trend, in  Fig.~\ref{fig:q-values} we plot the training time of federated PQoS as a function of~$\upsilon$. We consider  a scenario with $n=5$ vehicles and a training phase of $1\,500$ episodes. 
We see that the convergence time increases as $\upsilon$ decreases. In fact, the shorter periodicity implies more model updates within the training time, which in turn provides more data to the agent to optimize its decisions.
In particular, there is a significant gap between the systems using $\upsilon \geq 1.2$ s and $\upsilon \leq 1$ s. 
Based on the above considerations, we set $\upsilon=0.1$ s in the rest of the paper, so as to optimize the convergence.

%In the rest of the section, we analyze the performance in terms of the average reward $r$ obtained by each of the proposed solutions during the testing phase. Since all the learning strategies were trained to solve the same problem, we expect all of them to obtain similar performance. 
%At the same time, the complexity of the studied trade-off makes it possible for distinct approaches to converge to different solutions.
%Each solution may lead to a different balancing between \gls{qoe} and \gls{qos} even with a comparable average reward. 

%In figure below the following notation has been used: 
%\begin{itemize}
%    \item the number represents the number of user during the simulation;
%    \item \textbf{C} stands for centralized approach;
%    \item \textbf{D} stands for distributed approach;
%    \item \textbf{F} stands for federated approach.
%\end{itemize}
\begin{figure}[t!]
    \centering
    \setlength\belowcaptionskip{-0.5cm}
    \includegraphics[width=0.95\columnwidth]{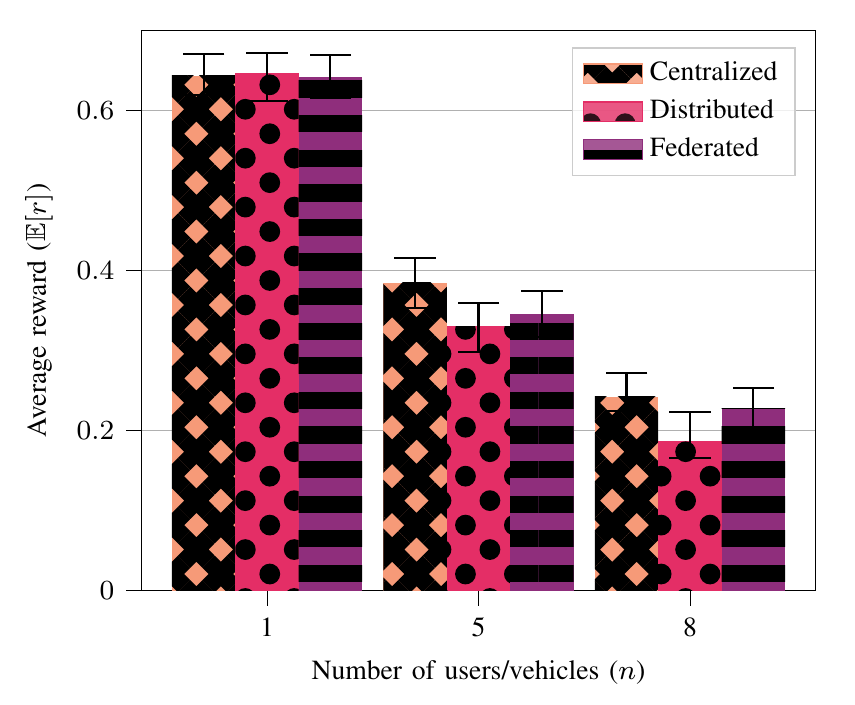}
    \caption{Average reward for different PQoS schemes (centralized, distributed, federated) vs. the number of users/vehicles $n$.}
    \label{fig:reward}
\end{figure}

In Fig.~\ref{fig:reward} we plot the average reward $\mathbb{E}[r]$, normalized in $[0,1]$ for better readability, considering centralized, distributed, and federated learning as a function of the number of vehicles $n$.
Notably, the reward is the main indicator of the system performance as it represents the trade-off between QoS and QoE of users/vehicles in the network. 
We can see that for $n=1$ the three schemes have the same average reward of around $0.65$. 
In this case, the network is underloaded, and is therefore able to serve most traffic requests regardless of the level of data compression at the application.
On the other hand, as $n$ increases, the network becomes more congested, which makes the choice of the optimal compression level non trivial. 
In this context, centralized PQoS outperforms any other competitor, and achieves a reward up to $13\%$ and $10\%$ bigger than distributed and federated PQoS, respectively.
In fact, RAN-AI leverages more input data to train the centralized \gls{rl} model, with respect to UE-AI in the decentralized scenario which is based only on local data.
For similar reasons, federated learning provides superior performance compared to distributed learning in all configurations, given that global model updates  can be used to improve the training of the local agents.
Notice that, for $n=8$, federated learning is only $6\%$ worse than centralized (vs. $22\%$ for distributed), and stands out as a desirable solution to support PQoS even in congested networks, while also promoting~privacy.
%As stated before, keeping constant the number of training episodes, and increasing the number of users, we are decreasing the input data for each \gls{ai}-model and therefore obtaining a lower reward.
%It can be immediately noticed the fact that, as expected, the fixed policies have defined behaviors. In particular, C-R obtains the value of \gls{qoe} equal to 1and reduces the QoS as the number of users grows, instead C-SA reaches higher values of \gls{qos} at the cost of setting aside the \gls{qoe}.

\begin{figure}[t!]
    \centering
    \includegraphics[width=0.95\columnwidth]{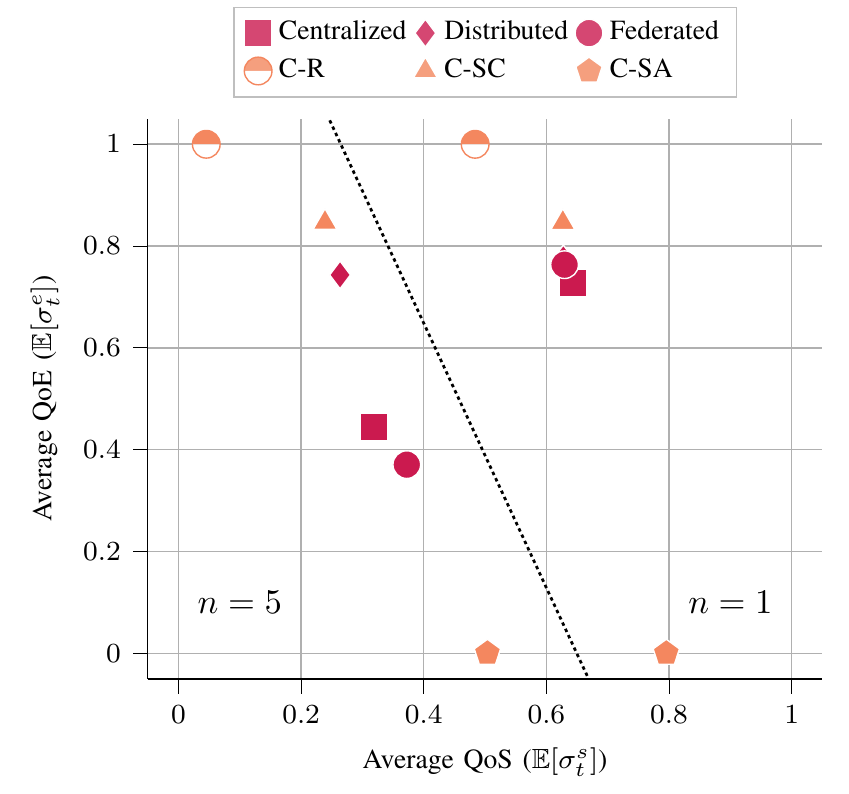}
    \caption{Average \gls{qos} vs. average \gls{qoe} for different PQoS schemes (centralized,
distributed, federated), compared to some constant baselines in which the compression level (C-R, S-SC, S-SA) is set a priori and does not change during the simulation. We consider $n=1$ and $n=5$.}
    \label{fig:big_comp}
\end{figure}

\begin{figure}[t!]
    \centering
    \setlength\belowcaptionskip{-0.5cm}
    \includegraphics[width=0.95\columnwidth]{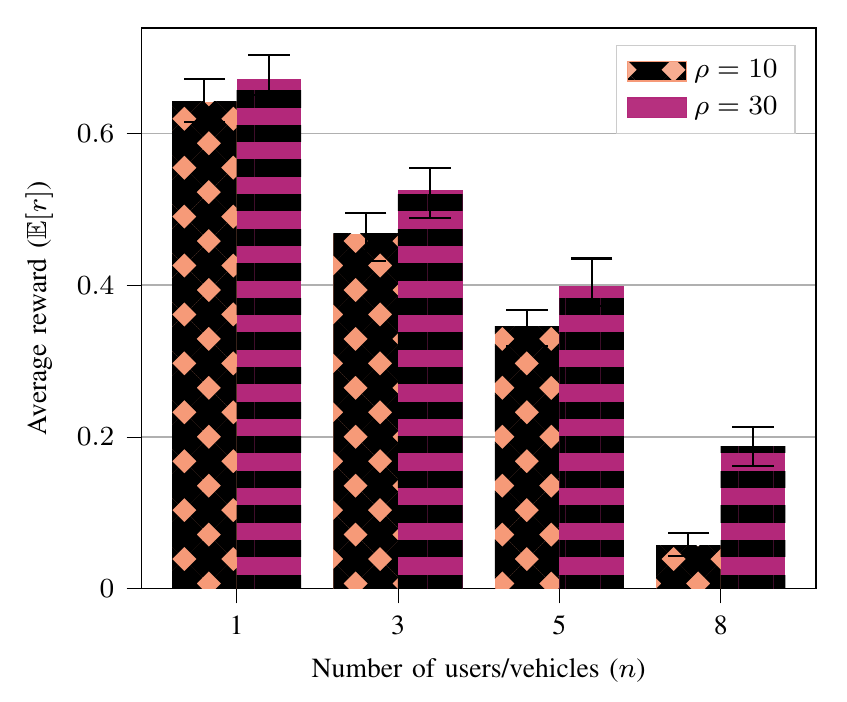}
    \caption{Average reward for federated PQoS, for different values of the penalty $\rho$ vs. the number of users/vehicles $n$.}
    \label{fig:reward-penalty}
\end{figure}

In Fig. \ref{fig:big_comp} we analyze the average \gls{qoe} and \gls{qos}, normalized in $[0,1]$, considering $n=1$ and $n=5$ vehicles, and vs. some constant benchmarks in which
the compression level (C-R, C-SA, C-SC) is set a priori and does not
change during the simulation.
Ideally, the optimal system will have $\mathbb{E}[\sigma_{t}^{s}]=1$ and $\mathbb{E}[\sigma_{t}^{e}]=1$, i.e., it will operate at the top-right corner of Fig. \ref{fig:big_comp}, to satisfy both QoS and QoE constraints, respectively.
As expected, \gls{cr} leads to the best performance in terms of \gls{qoe} given that data is not compressed before transmission, at the expense of a lower QoS even for $n=1$. In turn, \gls{csa} outperforms any other solution in terms of \gls{qos}, even though the QoE is close to zero, given that compression privileges efficiency over accuracy. 
On the other hand, PQoS can adapt the compression level to the conditions of the scenario via RL, and achieves the best trade-off between QoS and QoE.
Notably, PQoS discourages an overly aggressive behavior, and stands in the middle between the constant C-R and C-SA benchmarks. 
Notice that, as $n$ increases, distributed PQoS seems to prioritize \gls{qoe} over QoS, while centralized and federated PQoS have similar results. Specifically, even in congested networks, federated PQoS can improve the QoS by up to $60\%$ compared to constant C-SA, i.e., only 20\% less than C-SA which would in turn lead to a severe violation of the QoE.

Based on the above results, we identified federated learning as a good compromise for PQoS. Then, in Fig. \ref{fig:big_comp} we analyze the average reward of federated PQoS as a function of the penalty $\rho$ used to compute $\sigma_{t}^{e}$ in Eq.~\eqref{eq:qoe_value}. 
We observe that, while the \gls{qoe} is maximized using $\rho = 10$ as done in \cite{mason2022reinforcement}, $\rho = 30$ ensures a better trade-off between \gls{qos} and \gls{qoe}.
In particular, unlike $\rho=30$, when $\rho=10$ the reward drops from around $0.6$ for $n=1$ to less than $0.05$ for $n=8$.
This result demonstrates that, under federated learning, the reward function should be carefully selected;  specifically, $\rho$ should be increased as the number of users/vehicles increases to maintain the network stable.

%Concerning the three approaches, it is clear that they follow the same behavior when only one user is instantiated. Whenever the number of users increases, results tend to differ, with centralized approach keeping constant the ratio between \gls{qos} and \gls{qoe}, distributed having higher values of \gls{qoe} and federated that changes behaviour with 8 users, preferring to reduce the \gls{qos} to improve \gls{qoe}. This last result is particularly evident when is used a penalty of 10, and trying to solve that we decided to apply a higher penalty value which indeed achieved better 

\section{Conclusion and future research}
\label{sec:concl}
In this paper we explored the concept of PQoS for teleoperated driving. In particular, we studied whether moving intelligence from a centralized learning architecture to the local users can support better and faster optimization. To do so, starting from a centralized benchmark, we developed a new entity called ``UE-AI'' that, interacting with a custom agent via distributed or federated learning, can choose the optimal level of compression to send LiDAR data to satisfy QoS and QoE constraints. We proved that decentralized PQoS has many advantages, including inherent  privacy protection, fast convergence, and accuracy. 
Notably, federated learning offers a good trade-off between centralized and distributed learning, even in congested networks.

These results motivate further research efforts in this domain. For example, we will extend the UE-AI introducing new functionalities for federated learning, like split computing at the edge, as well as to increase security and privacy of PQoS operations.

%Unlike in a centralized pre-configured resource allocation approach, in which radio resources are scheduled by the gNB via scheduling grants, we study the feasibility of a decentralized algorithm based on ML in which each agent autonomously optimizes its channel selection pol- icy relying only on the gNB feedback, without prior ad hoc message exchange with the gNB itself

\section{Acknowledgment}
This work was partially supported by the European Union under the Italian National Recovery and Resilience Plan (NRRP) of NextGenerationEU, 
partnership on ``Telecommunications of the Future'' (PE0000001 - program “RESTART”).

\bibliographystyle{abbrv}
\bibliography{refs}
\end{document}